\documentclass{myczjphys}         
\begin{document}
\title{On some meaningful inner product for real Klein-Gordon fields with positive semi-definite norm}  
%
\authori{Frieder Kleefeld\,\footnote{E-mail: {\sf kleefeld@cfif.ist.utl.pt}\,, URL: {\sf http:/$\!$/cfif.ist.utl.pt/$\sim$kleefeld/}}\,\footnote{Present address: Doppler Institute for Mathematical Physics
and Applied Mathematics \& Nuclear Physics Institute, Czech Academy of Sciences, 250 68 \v{R}e\v{z} near Prague, Czech Republic.}}      \addressi{Centro de
F\'{\i}sica das Interac\c{c}\~{o}es Fundamentais (CFIF), Instituto Superior T\'{e}cnico,\\ Av.\ Rovisco Pais, 1049-001 Lisboa, Portugal}
\authorii{}     \addressii{}
\authoriii{}    \addressiii{}
\authoriv{}     \addressiv{}
\authorv{}      \addressv{}
\authorvi{}     \addressvi{}
%
\headauthor{Frieder Kleefeld}            
\headtitle{On some meaningful inner product for real Klein-Gordon fields with positive semi-definite norm}             
\lastevenhead{Frieder Kleefeld: On some meaningful inner product for real Klein-Gordon fields $\ldots$} 
\refnum{A}
\daterec{XXX}    
\issuenumber{0}  \year{2006}
\setcounter{page}{1}
\maketitle

\begin{abstract}
A simple derivation of a meaningful, manifestly covariant inner product for real Klein-Gordon (KG) fields with positive semi-definite norm is provided which turns out --- assuming a symmetric bilinear form --- to be the real-KG-field limit of the inner product for complex KG fields reviewed by A.\ Mostafazadeh and F.\ Zamani in December, 2003, and February, 2006 (quant-ph/0312078, quant-ph/0602151, quant-ph/0602161). It is explicitly shown that the positive semi-definite norm associated with the derived inner product for real KG fields measures the number of active positive and negative energy Fourier modes of the real KG field on the relativistic mass shell. The very existence of an inner product with positive semi-definite norm for the considered real, i.e.\ neutral, KG fields shows that the metric operator entering the inner product does not contain the charge-conjugation operator. This observation sheds some additional light on the meaning of the C operator in the CPT inner product of PT-symmetric Quantum Mechanics defined by C.M.\ Bender, D.C.\ Brody and H.F.\ Jones. 
\end{abstract}
\section{Introduction} 
In order to be able to set up a quantum theory with a probabilistic concept for some given Hamilton operator the very existence of an inner product with positive semi-definite norm on the space of functions being subject to the Hamilton operator is crucial. For a very long time the setting of Quantum Mechanics (QM) and later Quantum Field Theory (QFT) was confined to {\em strictly} Hermitian Hamilton operators, and accordingly emerged very early a well known inner product with positive semi-definite norm associated with the name Max Born \cite{Pais:1982we}. As a great surprise it was noticed very recently that the postulate of strict Hermiticity for differential operators representing observables in a Quantum Theory (QT), i.e. QM and QFT, with probability concept is much too tight (see e.g.\ Refs.\ \cite{Caliceti:1980,Bender:1998ke,Bender:1998gh,Fernandez:1998,Znojil:2001,Znojil:2004xw,PTworkshop:2003,Mostafazadeh:2004mx,Kleefeld:2002au,Kleefeld:2004jb,Kleefeld:2005hf,Bender:2004sv,Bender:2002vv,Bender:2005fc}). Inspired by some important conjecture by D.\ Bessis (and J.\ Zinn-Justin) of 1992 on the reality and positivity of spectra for manifestly non-Hermitian Hamiltonians it was C.M.~Bender and S.~Boettcher in 1997 \cite{Bender:1998ke} who proposed that the reality property of spectra of Hamilton operators is connected with the anti-unitary \cite{robnik1986} PT-symmetry of Hamilton operators, i.e.\ symmetry under simultaneous space (P, parity) reflection and time (T) reversal. The reality and boundness of the spectrum for eigenstates with unbroken PT-symmetry has been proven meanwhile rigorously for a general class of PT-symmetric Hamilton operators \cite{Dorey:2001uw}. Unfortunately it had become also clear that the PT operation cannot be used as a metric for a quantum mechanical inner product as the resulting pseudo-norm would be indefinite \cite{Znojil:2001,Japaridze:2001py,Kleefeld:2004jb,Mostafazadeh:2004mx}. This noted indefiniteness caused researchers working in the field to recall and review the long history of the concept of pseudo-Hermiticity  and indefinite metrics \cite{Kleefeld:2004jb,Znojil:2001} going back to names like P.A.M.\ Dirac, W.\ Pauli, W.\ Heisenberg, L.S.\ Pontrjagin, M.G.\ Kre\u{\i}n, $\ldots$ in order to find a suitable inner product with positive semi-definite norm for functions subject to PT-symmetric Hamilton operators. The identification of a so-called CPT-inner product with positive semi-definite norm by C.M. Bender, D.C.\ Brody and H.F.\ Jones \cite{Bender:2002vv} (see e.g.\ also discussions in Refs.\ \cite{Bender:2004sa,Bender:2004ej,Mostafazadeh:2003az,Samsonov:2005}) led then finally to a well defined pseudo-Hermitian QM for {\em non-Hermitian PT-symmetric} Hamilton operators. PT-symmetric QM has been extended most recently \cite{Bender:2004sa} even to PT-symmetric QFT, while recent attempts to extend non-Hermitian PT-symmetric QT to some QT for arbitrary non-Hermitian Hamilton operators are yet in progress (see e.g.\ Refs. \cite{Kleefeld:2005hf,Kleefeld:2004jb,Kleefeld:2003zj,Kleefeld:2003dx,Kleefeld:2002au} and references therein). The open question on the existence of some meaningful inner product in the context of such a future non-Hermitian QT is clearly again of crucial importance.

Besides this interesting new branch of physics it remained rather unclear whether the existence of an inner product with a positive norm is some consequence of using differential equations of first order in time like the non-relativistic Schr\"odinger equation or the relativistic Dirac equation, or whether the existence of such a product can be shown also in the context of differential equations of second order in time like the well known relativistic KG equation. In fact, the main --- naive --- text-book folklore says that the inner product of a charged KG field is related to the charge current density and hence the related norm should be indefinite, while consequently the norm of the real, i.e.\ neutral, KG field should be vanishing. On the contrary it had been argued e.g.\ in Refs.\ \cite{Kleefeld:2003zj,Kleefeld:2002au,Kleefeld:2004jb} that a consistent treatment of neutral KG fields leads to non-vanishing conserved current densities which are tightly related to the energy current density of a neutral KG field. The key idea to construct an inner product even for {\em neutral} KG fields with a {\em positive semi-definite norm} being worked out in the present manuscript is to subtract the respective current densities related to negative energy states of the KG field yielding a negative semi-definite norm from the respective current densities related to positive energy states of the KG field yielding a positive semi-definite norm. The resulting inner product for the neutral KG field with an overall positive semi-definite norm, which is derived here in a rather simple way, shows up --- assuming a symmetric bilinear form --- to be the real-KG-field limit of an inner product with positive semi-definite norm developed by A.\ Mostafazadeh \cite{Mostafazadeh:2002xa,Mostafazadeh:2003az} and reviewed in detail \cite{Mostafazadeh:2006ug} by A.\ Mostafazadeh and F.~Zamani (MZ) for {\em complex-valued} KG-fields. As pointed out by MZ \cite{Mostafazadeh:2006ug} their KG inner product reduces in a certain limit to already existing KG inner products derived and discussed in the literature e.g.\ by \cite{Wald:1994} R.M.\ Wald, J.J.\ Halliwell \& M.E.\ Ortiz, and R.P.\ Woodard, but is different --- to our understanding --- from Ref.~\cite{Sazdjian:1986qn}. Our present short paper has been on one hand partially motivated by finding an efficiently fast derivation of a meaningful inner product for real, i.e.\ neutral, KG fields with positive semi-definite norm, on the other hand is the present paper motivated by the desire to gain some clearer interpretation of the resulting inner product for {\em neutral} KG fields than the one provided on the basis of a global gauge symmetry by MZ in the context of {\em complex-valued} KG fields. The resulting very intuitive interpretation traces the way to a deeper understanding of the meaning of the ``intrinsic \cite{Bender:2004ej} parity operator'' C identified by C.M.\ Bender, D.C.\ Brody and H.F.\ Jones \cite{Bender:2002vv} entering the CPT inner product of PT-symmetric Quantum Mechanics, and to the development of meaningful inner product and probability concept to be used in the context of a QT based on non-Hermitian Hamiltonians without PT-symmetry.
\section{Derivation of an inner product for real KG fields with positive norm}
According to standard textbooks on QFT the neutral (acausal) Klein-Gordon (KG) field $\varphi(x)$ has a {\em real} mass $m$ and is {\em strictly real-valued}, i.e.\ $\varphi(x)=\overline{\varphi(x)}$. The space-time evolution of $\varphi(x)$ is characterized by the following action:
\begin{equation} {\cal S}[\varphi]\; = \; \int d^4 x \; {\cal L} (x) \; = \; \frac{1}{2} \left\{ \Big(\partial_\mu \varphi(x)\Big)\Big(\partial^\mu \varphi(x)\Big) - m^2 \,\varphi(x)^2 \right\} \; . \end{equation}
Throughout this paper we use the metric $g_{\mu\nu} = \mbox{Diag}(+,-,-,-)$. Lagrange's equation of motion resulting from $\delta {\cal S}[\varphi]=0$ is the well known KG equation:
\begin{equation} 0 \; = \; ( \partial^2 + m^2 )\,\varphi(x)  \; = \; \left( -\,i \partial_t + \sqrt{D}\, \right)\left( i \partial_t + \sqrt{D}\, \right)\,\varphi(x) \; , \label{kgequ1}
\end{equation}
with $D\equiv - \vec{\nabla}^2+ m^2$. It is well known that the solutions $\varphi^{(+)}(x)$ and $\varphi^{(-)}(x)$ of the underlying Schr\"odinger-like equations
\begin{equation} \left( -\,i \partial_t + \sqrt{D}\, \right) \varphi^{(+)}(x) \equiv 0 \qquad \makebox{and} \qquad \left( i \partial_t + \sqrt{D}\, \right) \varphi^{(-)}(x) \equiv 0 \label{schroeq1} \end{equation}
are simultaneously solutions of the original KG equation Eq.\ (\ref{kgequ1}) with positive and negative energy (Note, that $\sqrt{D}$ is strictly positive for $m>0$):
\begin{equation} 0 \; = \; ( \partial^2 + m^2 )\,\varphi^{(+)}(x) \quad , \quad  0 \; = \; ( \partial^2 + m^2 )\,\varphi^{(-)}(x) \, .
\end{equation}
For convenience we rewrite Eqs.\ (\ref{schroeq1}) as
\begin{equation} \varphi^{(+)}(x) \; = \; \frac{1}{\sqrt{D}} \;i \partial_t \, \varphi^{(+)}(x)  \; , \qquad  - \, \varphi^{(-)}(x) \; = \; \frac{1}{\sqrt{D}} \;i \partial_t \, \varphi^{(-)}(x) \; \label{speceq1} \end{equation}
yielding in particular also
\begin{equation} \varphi^{(+)}(x) - \varphi^{(-)}(x) \; = \; \frac{1}{\sqrt{D}} \;i \partial_t \, \Big( \varphi^{(+)}(x) + \varphi^{(-)}(x) \Big)  \; .\; \label{speceq2} \end{equation}
Now we make use of the well known fact that --- after defining $\omega(\vec{p}\,) \equiv \sqrt{\vec{p}^{\,2} + m^2}$ with $\omega(\vec{0})\equiv m$ --- any solution $\varphi(x)$ of the original KG equation Eq.\ (\ref{kgequ1}) can be decomposed as $\varphi(x) = \varphi^{(+)}(x) + \varphi^{(-)}(x)$:
\begin{eqnarray} \varphi(x) & = & \int \frac{d^4 p}{(2\pi)^3} \; \delta (p^2 - m^2) \; e^{-i\,p\cdot x} \; a(p)  \nonumber \\
 & = & \int \frac{d^3 p}{(2\pi)^3} \, \int dp_0 \; \delta \left(\,p_0^2 - \omega(\vec{p}\,)^2\right) \; e^{-i\,p\cdot x} \; \Big(\theta(p_0)\;  a(p) + \theta(-p_0)\;  a(p) \Big) \nonumber \\
 & = & \int \frac{d^3 p}{(2\pi)^3\,2\,\omega(\vec{p}\,)} \, \Big(e^{-i\,p\cdot x} \;  a(p) + e^{+i\,p\cdot x} \; a(-p) \Big)\Big|_{p^0=\omega(\vec{p}\,)} \nonumber \\
 & \stackrel{!}{=} & \int \frac{d^3 p}{(2\pi)^3\,2\,\omega(\vec{p}\,)} \, \Big(e^{-i\,p\cdot x} \;  a(\vec{p}\,) + e^{+i\,p\cdot x} \; \overline{a(\vec{p}\,)} \Big)\Big|_{p^0=\omega(\vec{p}\,)} \nonumber \\[1mm]
 & = &  \varphi^{(+)}(x) + \varphi^{(-)}(x) \; , 
\end{eqnarray}
with $a(\vec{p}\,)\equiv a(p)|_{p^0=\omega(\vec{p}\,)}$ and $\overline{a(p)}\stackrel{!}{=} a(-p)$ due to $\varphi(x)=\overline{\varphi(x)}$ implying of course $\varphi^{(+)}(x) \stackrel{!}{=} \overline{\varphi^{(-)}(x)}$. The decomposition $\varphi(x) = \varphi^{(+)}(x) + \varphi^{(-)}(x)$ in combination with Eqs.\ (\ref{speceq1}), (\ref{speceq2}) and (\ref{kgequ1}) yields obviously the following important relations:
\begin{eqnarray} \varphi^{(\pm)}(x) & = & \frac{1}{2} \Big(\varphi^{(+)}(x) + \varphi^{(-)}(x) \Big) \pm \frac{1}{2} \Big(\varphi^{(+)}(x) - \varphi^{(-)}(x) \Big) \nonumber \\[2mm]
 & \stackrel{!}{=} & \frac{1}{2} \left( 1 \pm \frac{1}{\sqrt{D}} \;i \partial_t \right)\,\varphi(x) \; , \label{speceq5}\\[2mm]
i \partial_t \,\varphi^{(\pm)}(x)  & = & \frac{1}{2} \left( i \partial_t \mp \frac{1}{\sqrt{D}} \;\partial^2_t \right)\,\varphi(x) \; = \; \frac{1}{2} \left( i \partial_t \pm \frac{1}{\sqrt{D}} \;D \right)\,\varphi(x) \nonumber \\[2mm]
  & = & \frac{1}{2} \left( i \partial_t \pm \sqrt{D} \right)\,\varphi(x) \; .\label{speceq6}
\end{eqnarray}
At this place it is worth mentioning that --- as we are dealing here with {\em real}, i.e.\ {\em neutral}, KG fields --- contrary to what is discussed in Ref.\ \cite{Mostafazadeh:2003az} \footnote{The author of Ref.\ \cite{Mostafazadeh:2003az} writes in the context of Eq.\ (51) of Ref.\ \cite{Mostafazadeh:2003az}: {\em ``$\ldots$ This is sufficient to infer that indeed ${\cal H}={\cal H}_+ \oplus {\cal H}_-$. The generalized charge-conjugation operator C $\ldots$ is actually the grading operator associated with this orthogonal direct sum decomposition of ${\cal H}$, i.e., if $\psi=\psi_+ + \psi_-$ such that $\psi_\pm\in{\cal H}_\pm$, then $C\,\psi = \psi_+ - \psi_-$. In other words, C is the operator that decomposes the Hilbert space into its positive- and negative-energy subspaces. In view of the fact that for a complex Klein-Gordon field the positive and negative energies respectively occur for positive and negative charges, C is identical with the ordinary charge-conjugation operator $\ldots$''.}} the operator $(1/\sqrt{D}) \, i\partial_t$ appearing in our Eq.\ (\ref{speceq2}), i.e.\ $\varphi^{(+)}(x) - \varphi^{(-)}(x)=(1/\sqrt{D}) \, i\partial_t \,\varphi(x)$, {\em cannot} have the interpretation of a {\em charge-conjugation} operator typically denoted as C, while it --- in accordance to the conclusions at the end of this manuscript --- behaves in many aspects like a number operator.
For the definition of a suitable scalar product we have to consider now two {\em real-valued} KG fields $\varphi_1(x) = \varphi^{(+)}_1(x) + \varphi^{(-)}_1(x)$ and $\varphi_2(x) = \varphi^{(+)}_2(x) + \varphi^{(-)}_2(x)$ of equal {\em real-valued} mass $m$ respecting not only the KG equations $(\partial^2 + m^2)\, \varphi_1(x)=0$ and $(\partial^2 + m^2)\, \varphi_2(x)=0$, yet also:
\begin{equation} (\partial^2 + m^2)\,\varphi^{(\mp)}_1(x) = 0\; , \quad
 (\partial^2 + m^2)\,\varphi^{(\pm)}_2(x) =  0 \; .
\end{equation}
After multiplying the left-hand and right-hand equations with $\varphi^{(\pm)}_2(x)$ and $\varphi^{(\mp)}_1(x)$, respectively, we subtract the left-hand from the right-hand equations to obtain  continuity equations which had been investigated and interpreted e.g.\ in Refs.\ \cite{Kleefeld:2003zj,Kleefeld:2002au,Kleefeld:2004jb} ($\stackrel{\leftrightarrow}{\partial_\mu}\equiv \partial_\mu - \stackrel{\leftarrow}{\partial_\mu}$):
\begin{equation} \partial_\mu \Big( \varphi^{(-)}_1(x) \, \stackrel{\leftrightarrow}{\partial^\mu}\, \varphi^{(+)}_2(x)  \Big)\; = \; 0 \; , \qquad
\partial_\mu \Big( \varphi^{(+)}_1(x) \, \stackrel{\leftrightarrow}{\partial^\mu}\, \varphi^{(-)}_2(x) \Big) \; = \; 0 \; .\end{equation}
Any linear combination of the two continuity equations leads again to a continuity equation. Hence, we construct for later convenience the conserved current density
\begin{eqnarray} \lefteqn{j_{\varphi_1\varphi_2}^\mu(x) \; =}\nonumber \\
 & = & b\,\left\{ (a+1)\, \Big( \varphi^{(-)}_1(x) \, i\,\stackrel{\leftrightarrow}{\partial^\mu}\, \varphi^{(+)}_2(x) \Big) +  (a-1)\,  \Big( \varphi^{(+)}_1(x) \, i\, \stackrel{\leftrightarrow}{\partial^\mu}\, \varphi^{(-)}_2(x)  \Big) \right\} \; \nonumber \\ 
 & = &  \frac{b}{2}\, \Bigg\{ \left( \frac{1}{\sqrt{D}}\; \partial_t \, \varphi_1(x) \right)  \stackrel{\leftrightarrow}{\partial^\mu} \varphi_2(x) - \,\varphi_1(x) \stackrel{\leftrightarrow}{\partial^\mu}\left( \frac{1}{\sqrt{D}}\; \partial_t \, \varphi_2(x) \right) \nonumber \\
 & &\quad  + i a \,\left( \varphi_1(x) \stackrel{\leftrightarrow}{\partial^\mu} \varphi_2(x) + \left( \frac{1}{\sqrt{D}}\; \partial_t \, \varphi_1(x) \right)\stackrel{\leftrightarrow}{\partial^\mu}\left( \frac{1}{\sqrt{D}}\; \partial_t \, \varphi_2(x) \right)\right) 
\Bigg\} \; ,
\label{speceq10}
\end{eqnarray}
where the constants $a$ and $b$ are not yet specified and the right-hand side of the identity was obtained by applying Eqs.\ (\ref{speceq5}) and (\ref{speceq6}). Similarly  the respective conserved charge providing a {\em bilinear} form for {\em real-valued} KG fields $\varphi_1(x)$ and $\varphi_2(x)$ can be reexpressed with the help of Eqs.\ (\ref{speceq5}) and (\ref{speceq6}) in the following way:
\begin{eqnarray} \lefteqn{(\varphi_1,\varphi_2)_{b,a} \; \equiv \; \int d^3 x \; j_{\varphi_1\varphi_2}^0(x)\;=} \nonumber \\
 & = & \int d^3 x \;\, b \, \Big\{ (a+1)\, \Big( \varphi^{(-)}_1(x) \, i\stackrel{\leftrightarrow}{\partial_t} \varphi^{(+)}_2(x) \Big)  +  (a-1)\,  \Big( \varphi^{(+)}_1(x) \, i\stackrel{\leftrightarrow}{\partial_t} \varphi^{(-)}_2(x) \Big) \Big\} \nonumber \\
 & = & \int d^3 x \;\, \frac{b}{2} \, \Big\{  - \Big( i\,\partial_t \,\varphi_1(x) \Big)\, \frac{1}{\sqrt{D}} \;i\,\partial_t \, \varphi_2(x) - \,\left( \frac{1}{\sqrt{D}} \;i\,\partial_t \, \varphi_1(x) \right)\Big( i\,\partial_t \,\varphi_2(x) \Big) \nonumber  \\
 & &  \qquad\qquad + \,\varphi_1(x) \, \sqrt{D}\;  \varphi_2(x) + \, \Big(\sqrt{D}\;  \varphi_1(x) \Big)\; \varphi_2(x) + \, i a  \, \Big[ \varphi_1(x) \stackrel{\leftrightarrow}{\partial_t} \varphi_2(x) \nonumber  \\
 &  & \qquad\qquad + \Big( \sqrt{D} \;\varphi_1(x) \Big) \, \frac{1}{\sqrt{D}} \;\partial_t \, \varphi_2(x) - \,\left( \frac{1}{\sqrt{D}} \;\partial_t \, \varphi_1(x)\right) \Big( \sqrt{D} \;\varphi_2(x) \Big) \Big] \Big\} \nonumber \\
 & \stackrel{!}{=} & \int d^3 x \; b \, \Big\{ \varphi_1(x) \, \sqrt{D}\,  \varphi_2(x)  +  \Big( \partial_t \,\varphi_1(x) \Big) \frac{1}{\sqrt{D}} \;\partial_t \, \varphi_2(x)  + \,i a \, \varphi_1(x) \stackrel{\leftrightarrow}{\partial_t} \varphi_2(x) \Big\}  \nonumber \\
 & \stackrel{!}{=} & \int d^3 x \; b \; \Big( D^{1/4} \,\varphi_1(x) , \, D^{-1/4} \, \partial_t \,\varphi_1(x)\Big) \underbrace{\left( \begin{array}{cc} 1 & i\,a \\[2mm] -i\,a & 1 \end{array} \right)}_{\equiv \cal M}  \left( \begin{array}{c} D^{1/4} \,\varphi_2(x) \\[2mm] D^{-1/4} \, \partial_t \,\varphi_2(x)  \end{array}\right) . \nonumber \\
\label{speceq7}
\end{eqnarray}
In the last two steps we applied partial integrations. Our result for the bilinear form being applicable to {\em real-valued} KG fields coincides --- after replacing $\varphi_1(x)$ by $\overline{\varphi_1(x)}$ --- with the KG inner product in Eq.\ (12) of the first paper of Ref.\ \cite{Mostafazadeh:2006ug}, where {\em complex-valued} KG fields had been assumed with $b\in I\!\!R^{\,+}$ and the condition $a\in [-1,+1]$ to keep the eigenvalues $1\pm a$ of the matrix ${\cal M}$ real and non-negative. For the here considered case of {\em real-valued} KG fields we have to confine ourselves to a {\em symmetric} matrix ${\cal M}={\cal M}^T$ yielding $a\equiv 0$ in order to achieve a {\em real} and {\em symmetric} bilinear form $(\varphi_1,\varphi_2)_{b,0}=(\varphi_2,\varphi_1)_{b,0}$ possessing all properties of an inner product\footnote{The current densitiy takes for $a=0$ the simple form $j_{\varphi_1\varphi_2}^\mu(x) = 2\, b \; \mbox{Re} [ \, \varphi^{(-)}_1(x) \, i\stackrel{\leftrightarrow}{\partial^\mu}\, \varphi^{(+)}_2(x) \, ]$ being obviously real for $b\in I\!\!R^{\,+}$.}. This confirms what has been claimed in Ref.\ \cite{Mostafazadeh:2006ug} that the relativistically invariant positive-definite inner product on the space of {\em real} KG fields is unique, {\em if} --- as we would like to add --- other, manifestly different constructions for KG inner products like the one in Ref.\ \cite{Sazdjian:1986qn} can be excluded. Inspection of Eq.\ (\ref{speceq7}) implies that the norm $||\varphi||_{b}^{\,2}\equiv(\varphi,\varphi)_{b,a}$ and the underlying current density $j_{\varphi\varphi}^\mu(x)$ of the {\em real} KG field are independent of the paramter $a$ even for $a\not=0$:
\begin{eqnarray} j_{\varphi\varphi}^\mu(x) 
 & = &  b \; \left( \frac{1}{\sqrt{D}}\; \partial_t \, \varphi(x) \right)  \stackrel{\leftrightarrow}{\partial^\mu} \varphi(x) \; ,\nonumber \\[1mm]
 ||\varphi||_{b}^{\,2} & = &  \int d^3 x \;\, b \, \Big\{ \varphi(x) \, \sqrt{D}\,  \varphi(x)  +  \Big( \partial_t \,\varphi(x) \Big) \frac{1}{\sqrt{D}} \;\partial_t \, \varphi(x) \Big\} .
\end{eqnarray}
\section{Interpretation of the derived inner product for real KG fields}
To gain some interpretation of the conserved the inner product $(\varphi_1,\varphi_2)_{b,0}$ and the respective norm $||\varphi||_{b}^{\,2}$ we recall that there holds,
\begin{eqnarray} \varphi_1^{(+)}(x) & = & \int \frac{d^3 p}{(2\pi)^3\,2\,\omega(\vec{p}\,)} \;\,e^{-\,i\,(\omega(\vec{p}\,)\, t-\,\vec{p}\cdot \vec{x})}\;  a_1(\vec{p}\,) \;, \nonumber \\
 \varphi_2^{(+)}(x) & = & \int \frac{d^3 p}{(2\pi)^3\,2\,\omega(\vec{p}\,)} \;\,e^{-\,i\,(\omega(\vec{p}\,)\, t-\,\vec{p}\cdot \vec{x})}\;  a_2(\vec{p}\,) \;, \nonumber \\
 \varphi_1^{(-)}(x) & = & \int \frac{d^3 p}{(2\pi)^3\,2\,\omega(\vec{p}\,)}\; \,e^{+\,i\,(\omega(\vec{p}\,)\, t-\,\vec{p}\cdot \vec{x})}\;  \overline{a_1(\vec{p}\,)} \;, \nonumber \\
 \varphi_2^{(-)}(x) & = & \int \frac{d^3 p}{(2\pi)^3\,2\,\omega(\vec{p}\,)} \;\,e^{+\,i\,(\omega(\vec{p}\,)\, t-\,\vec{p}\cdot \vec{x})}\;  \overline{a_2(\vec{p}\,)} \;,
\end{eqnarray}
yielding due to Eq.\ (\ref{speceq7}) obviously
\begin{eqnarray} (\varphi_1,\varphi_2)_{b,0} & = & \int d^3 x \,  \int \frac{d^3 p_1}{(2\pi)^3\,2\,\omega(\vec{p}_1)}\int \frac{d^3 p_2}{(2\pi)^3\,2\,\omega(\vec{p}_2)}\;\; b \;\Big( \omega(\vec{p}_2) + \omega(\vec{p}_1) \Big) \,\times \nonumber \\[1mm]
 &  & \quad \times \, \Big\{ \overline{a_1(\vec{p}_1)} \; a_2(\vec{p}_2)\;e^{-\,i\,(\vec{p}_1-\vec{p}_2)\cdot \vec{x}} +   \;a_1(\vec{p}_1) \;\overline{a_2(\vec{p}_2)}\;e^{+\,i\,(\vec{p}_1-\vec{p}_2)\cdot \vec{x}}\Big\} \nonumber \\
 & = &  \int \frac{d^3 p}{(2\pi)^3\,2\,\omega(\vec{p})}\;\; b \; \Big\{ \overline{a_1(\vec{p})} \; a_2(\vec{p}) + \;a_1(\vec{p}) \;\overline{a_2(\vec{p})}\Big\} \nonumber \\
 & = &  \int \frac{d^3 p}{(2\pi)^3\,2\,\omega(\vec{p})}\;\; 2\, b \; \mbox{Re} \, \Big[\,\overline{a_1(\vec{p})} \; a_2(\vec{p})\,\Big] \; .
\end{eqnarray}
With $a_1(\vec{p}\,)\equiv |a_1(\vec{p}\,)|\,\exp(i \,\chi_1(\vec{p}\,))$ and $a_2(\vec{p}\,)\equiv |a_2(\vec{p}\,)|\,\exp(i \,\chi_2(\vec{p}\,))$ we obtain therefore
\begin{eqnarray}(\varphi_1,\varphi_2)_{b,0} & = & \int \frac{d^3 p}{(2\pi)^3\,2\,\omega(\vec{p})}\;\; 2\, b \; |a_1(\vec{p}\,)|\,|a_2(\vec{p}\,)|\,\cos\Big(\chi_1(\vec{p})-\chi_2(\vec{p})\Big) \; , \nonumber \\
||\varphi||_{b}^{\,2} & = & b \int \frac{d^3 p}{(2\pi)^3\,2\,\omega(\vec{p})}\;\; 2\;  |a(\vec{p}\,)|^2 \; \stackrel{!}{=} \;b \int \frac{d^4 p}{(2\pi)^3} \; \delta (p^2 - m^2) \; |a(p)|^2  . \;\;\;\;
\end{eqnarray}
Hence, the norm $||\varphi||_{b}^{\,2}$ being real and positive for $b\in I\!\!R^{\,+}$ measures the Lorentz-invariant number of active Fourier modes of the real KG field on the relativistic mass shell $p^2=m^2$ taking into account positive {\em and} negative energies.

In view of the foregoing results it is certainly very tempting to choose $b=1$.

\bigskip
{\small \indent This work has been supported by the
{\em Funda\c{c}\~{a}o para a Ci\^{e}ncia e a Tecnologia} \/(FCT) of the {\em Minist\'{e}rio da Ci\^{e}ncia e da Tecnologia e do Ensino Superior} \/of Portugal, under grants no.\ PRAXIS XXI/BPD/20186/99, SFRH/BDP/9480/2002, POCTI/FNU/49555/2002, POCTI/FNU/50328/2003, POCTI/FP/FNU/63437/2005, and the Czech project LC06002.}

\end {document}